\documentstyle[preprint,aps,epsf,floats,tighten,eqsecnum]{revtex}

\begin{document}

\preprint{
\vbox{\hbox{JHU--TIPAC--98011}
      \hbox{hep-ph/9811405}
      \hbox{November 1998} }}

\title{Radiative Leptonic $B_c$ Decays in Effective Field
Theory}
\author{George Chiladze, Adam F.~Falk and Alexey A.~Petrov}
\address{Department of Physics and Astronomy, The Johns Hopkins
University\\ 3400 North Charles Street, Baltimore, Maryland
21218}
\maketitle
\thispagestyle{empty}
\setcounter{page}{0}
\begin{abstract}%
The recent discovery of the $B_c$ meson by the
CDF collaboration and proposed new experiments at Fermilab and
CERN motivate new theoretical studies of the $B_c$ system.
Here we investigate the radiative leptonic decay
$B_c\to\gamma l\nu$.  This process is an important
background to the annihilation process $B_c\to l\nu$, which
will be used to extract the $B_c$ decay constant.  We perform
a model-independent calculation, based on QCD, of the
partial width and various kinematic distributions.  We also
examine the decay within the framework of NRQCD, an effective
field theory of nonrelativistic quarks, generalizing the NRQCD
Lagrangian to include external sources for the weak and
electromagnetic currents.  Finally, we will show how NRQCD
reproduces the correct position of the $B_c^*$ pole in the
emission of very soft photons.
\end{abstract}

\newpage

\section{Introduction}

The $B_c$ is probably the final long lived pseudoscalar meson
which will be found in our lifetimes.  Its recent discovery at
CDF~\cite{CDFBc}, and the future prospect of thorough
experimental studies at LHC and BTeV, motivates extensive
theoretical studies of this system~\cite{Ger95}.  In
particular, the $B_c$ offers a unique possibility to study the
effects of weak interactions in a quarkonium-like environment.
Since it is composed of heavy quarks of two different
flavors, the $B_c$ is stable against strong annihilation decays
and its dynamics approaches a simple perturbative limit as
$m_c,m_b\to\infty$. In this limit, the $B_c$ is a very compact
bound state of a $c$ and $\bar b$ quark, with a small
admixture of non-perturbative higher Fock states containing
gauge bosons and light quark-antiquark pairs. This admixture
is small because soft nonperturbative gluons, with large
Compton wavelengths, have little overlap with the compact
$B_c$ state. It is useful to study the $B_c$ system, like
quarkonium, in the framework of an effective nonrelativistic
quantum field theory~\cite{bbl}.

The advantages of an effective field theory approach are both
conceptual and quantitative.  While one would intuit that the
$m_c,m_b\to\infty$ limit is one in which a description based
on a constituent quark model should work reasonably well, an
effective field theory based on the operator product
expansion puts this intuition on a rigorous basis.  It is
unsurprising, of course, that the results we obtain in this
paper are similar in structure to those of the quark model,
since the quark model does respect the symmetries of the
heavy quark limit.  What an effective field theory does, that
the quark model does not, is provide an oganized expansion
within which the leading corrections to this limit may be
accomodated.  While we will not compute higher order
corrections in this paper, we will use the power counting of
the effective field theory to estimate their size, thereby
casting light on the accuracy of our results.

One basic characteristic of the $B_c$ meson is its decay
constant, defined by
\begin{equation}\label{decayconst}
  \langle 0|\,\bar b\gamma^\mu\gamma^5c\,|B_c(p) \rangle =
  if_{B_c}p^\mu\,.
\end{equation}  This is a QCD definition, although in quark
models and in the nonrelativistic limit $f_{B_c}$ may be
identified with the value of the wave function at the origin.
The decay constant probes the strong QCD dynamics which is
responsible for the binding of the quark-antiquark state. The
most straightforward way to determine $f_{B_c}$ would be to
measure the purely leptonic partial widths, such as
$B_c \to l \nu_l$, where $l=e,\mu,\tau$ \cite{BrF95}.  The
rate for this process is given by
\begin{equation}
  \Gamma (B_c \to l \nu) =
  \frac{G_F^2}{8 \pi} |V_{cb}|^2 f_{B_c}^2 m_{B_c}^3
  \frac{m_l^2}{m_{B_c}^2}
  \left(1-\frac{m_l^2}{m_{B_c}^2} \right)^2.
\end{equation} However, the practical usefulness of this method
is limited, because the decaying meson is spinless and this
mode is helicity suppressed.  A helicity flip on an external
lepton line is required, leading to a suppression of the rate
by an additional factor $m_l^2/m_{B_c}^2$. This suppression is
$6\times10^{-8}$ for $B_c\to e\nu_e$, and $3\times10^{-4}$ for
$B_c\to\mu\nu_\mu$.  The only leptonic mode with a substantial
branching fraction is $B_c\to\tau\nu_\tau$, but this is
difficult to observe because the $\tau$ must be reconstructed
from its decay products.

The helicity suppression can be overcome if there is third
particle in the final state. In particular, adding a photon
does not change the fact that the decay rate is proportional
to the decay constant.  A na\"\i ve estimate suggests
that for $l=\mu$ the additional electromagnetic coupling is
effectively compensated by the lifting of the helicity
suppression, since
$(\alpha/4 \pi)/(m_{B_c}^2/m_\mu^2) \sim 2$.  For $l=e$, the
same counting suggests that the radiative leptonic mode
dominates the purely leptonic decay.  If branching ratios as
small as that for $B_c\to\mu\nu_\mu$ are eventually measured,
then there are two consequences.  First,
$B_c\to\gamma\mu\nu_\mu$ will be an important background (or
tool) for the extraction of $f_{B_c}$.  Second,
$B_c\to\gamma e\nu_e$ will be observable even though
$B_c\to e\nu_e$ is not.  For both of these reasons, it is
important to understand the radiative decay process.

Radiative leptonic $B_c$ decays already have been studied using
quark potential models~\cite{Cha97} and QCD sum
rules~\cite{Ali98}.  We will comment on these approaches later
and relate them to our own, which will be based on QCD and
its nonrelativistic expansion, NRQCD.  The corresponding decay
for systems with one heavy and one light quark,
$B\to\gamma l\nu_l$, also has been examined~\cite{Bur95},
employing insofar as possible a heavy fermion expansion for the
$b$ quark.  Unfortunately, that decay is dominated by photons
radiated from the $u$, the effect of which is impossible to
compute model-independently.  The advantage of the $B_c$ system
is that it can be treated systematically in an expansion in
${\bf v}$, the nonrelativistic three-velocity of the $\bar b$
and the $c$ in their mutual bound state.  The expansion may
be used to justify the application of perturbative QCD to
this decay~\cite{bbl}.  In addition, we note that, in contrast
to $B\to l\nu,\gamma l\nu$, the decays $B_c\to l\nu,\gamma
l\nu$ are not CKM-suppressed.

The value of NRQCD is that it provides a rigorous counting of
powers of ${\bf v}$, which may be applied to expectation
values of heavy quark operators in  external states $Q\overline
Q'$~\cite{bbl} where $m_Q,m_{Q'}\gg\Lambda_{\rm QCD}$.  This
power counting is related to the heavy quark effective theory
(HQET) derivative expansion in powers of $D_\mu/m_Q$, although
it differs in some details.  What NRQCD and HQET share is a
``spin symmetry'', by which the magnetic interactions of the
heavy quarks decouple as $1/m_Q$.  In general, the power of
${\bf v}$ which NRQCD assigns to an operator depends on the
matrix element which is being taken.  Therefore the NRQCD
expansion must be performed on the matrix elements, not just
on the operators.  By contrast, the HQET power counting is
identical for all bound states to which it is applied, which
are of the form $Q\bar q$, where $m_q\alt\Lambda_{\rm QCD}$. 
Although the HQET expansion is technically also an expansion
of matrix elements, there is therefore no confusion in
thinking of it equally as applying to the operators themselves.

In fact, we will be able to treat our application of NRQCD
to the $B_c$ system as an operator expansion as well, because
of a number of simplifications which apply to the process under
consideration.  First, we will work only to leading order in
the NRQCD expansion, with no dependence on the $|\bar
bcg\rangle$ higher Fock components for which the power
counting in ${\bf v}$ can be subtle (and, of course,
interesting).  Second, the only external states which will be
important are the $B_c$ and the $B_c^*$, both of which are
dominated by the same spatial configuration of the $\bar bc$
pair. Therfore, the relevant leading matrix elaments are
related by heavy quark spin symmetry. The only matrix element
which will appear is the one which defines
$f_{B_c}$ in NRQCD, namely
\begin{equation}
  im_{B_c} f_{B_c} = \langle 0 | \chi_b^\dagger \psi_c |B_c
  \rangle\,,
\end{equation} where $\chi_b$ and $\psi_c$ are the
two-component NRQCD field operators.  Third, the final state
consists of no particles which are strongly interacting.  As a
result, the operator product expansion, which NRQCD helps to
organize in the case of hadronic decays, is here a trivial
affair.

We will first study the decay $B_c\to l\nu\gamma$ in
perturbative QCD, which is fairly simple since there are no
infrared divergences or other subtleties.  We will then match
the QCD result onto NRQCD and identify the leading operator
which contributes to the annihilation of the $B_c$. The
expansion turns out to be straightforward, but it is useful to
see how the parton model result emerges in the nonrelativistic
limit.  Finally, we will show how NRQCD reproduces the correct
position of the $B_c^*$ pole in the emission of very soft
photons.

\section{The decay in perturbative QCD}

\begin{figure}
\centerline{
\epsfysize 3in
\epsfbox{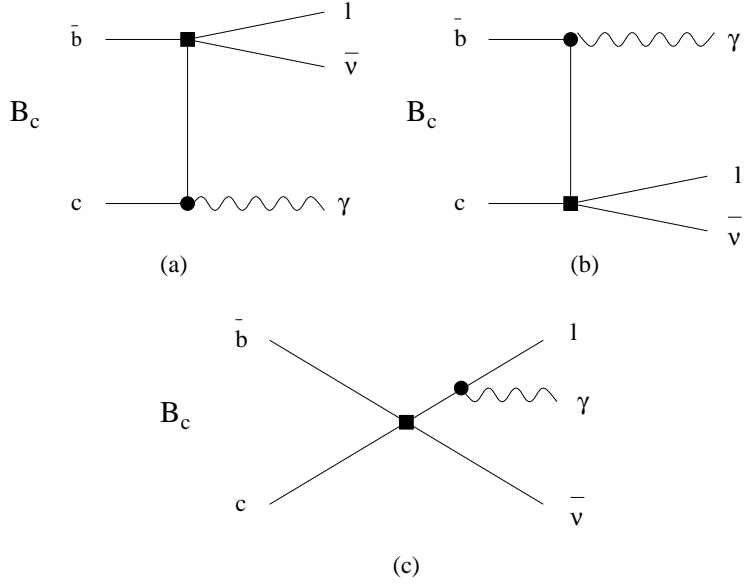}}
\vskip0.5cm
\caption{Feynman diagrams for $\bar b c\to\gamma l\nu_l$}
\label{QCDgraphs}
\end{figure}

We begin by calculating the rate for $B_c \to l \nu \gamma$
perturbatively in full QCD.  Since there are no strongly
interacting particles in the final state, the QCD operator
product expansion is trivial.  The matrix element for $B_c \to
l \nu \gamma$ is governed by the matrix element for $\bar b c
\to l \nu_l\gamma$, followed by the projection of the $\bar b
c$ onto the $B_c$ state.  At leading order, we take the quarks
to be in the (leading) $^1S_0$ Fock configuration. 

The computation is simple, involving the set of diagrams
presented in Fig.~\ref{QCDgraphs}. The decay amplitude can be
written as
\begin{eqnarray} \label{qcd1} 
  && {\cal A}(\bar b c[^1S_0] \to l
  \nu \gamma) = V_{cb}\frac{G_F}{\sqrt{2}}
  \epsilon^*_\nu \bar u(p_\nu) \gamma_\mu (1 -\gamma_5)
  v(p_l)\,\langle 0|\,\bar b  \gamma_\alpha (1 -\gamma_5)
  c\,|\bar b c \rangle\\ &&
  \qquad\qquad\qquad\times\Biggl\{
  \left[\frac{eQ_c p^\nu_c}{p_ck} - \frac{eQ_b p^\nu_b}{p_b k}
  +
  \frac{eQ_l p^\nu_l}{p_l k}\right] g^{\mu \alpha}
  - \left[\frac{eQ_c}{p_c k} \Gamma^{\mu \nu \alpha}
  - \frac{eQ_b p^\nu_c}{p_b k} \Gamma^{\nu \mu \alpha} -
  \frac{eQ_l p^\nu_c}{p_l k} \Gamma^{\alpha \nu \mu}
  \right]\Biggr \},\nonumber 
\end{eqnarray}  where
\begin{equation}
  \Gamma^{\mu \nu \alpha} = k^\mu g^{\nu\alpha} + k^\nu
  g^{\mu\alpha} - k^\alpha g^{\mu \nu} 
  - i \epsilon^{\mu \nu\alpha\beta} k_\beta.
\end{equation} Here $\epsilon^\mu$ and $k^\mu$ are the
polarization and momentum of the photon.  We now anticipate
the results of the next section, in which we will verify that
the current $\bar b\gamma_\alpha(1-\gamma_5)c$ is the leading
contribution to the decay, to justify the evaluation of the
$B_c$ matrix element (\ref{decayconst}).  Defining the leptonic
matrix element
$l_\mu=\bar u\gamma_\mu (1 - \gamma_5) v$, we find 
\begin{eqnarray} 
  {\cal A}(B_c \to l \nu \gamma) &=& -i
  V_{cb}
  \frac{G_F}{\sqrt{2}}\,f_{B_c}m_{B_c}l_\mu\epsilon^*_\nu 
   \Bigg\{\frac{e Q_l}{p_l k} v^\mu p_l^\nu
   +\left[\frac{eQ_c p^\nu_c}{p_ck} - \frac{eQ_b p^\nu_b}{p_b
   k}\right] v^\mu
    \nonumber \\ 
   && \qquad\qquad
   \mbox{}+ B_1(k v\, g^{\mu \nu} -k^\mu v^\nu) 
   - B_2\,i \epsilon^{\mu \alpha \nu \beta} 
  v_\alpha k_\beta \Bigg\} \,,
\end{eqnarray}  where  $v^\mu$ is the four-velocity of
the $B_c$,
$B_1 = eQ_c/2p_c k - eQ_b/2p_b k  + eQ_l/2p_l k$, and $B_2 =
eQ_c/2p_c k + eQ_b/2p_b k + eQ_l/2p_l k$.  This expression
simplifies somewhat if we choose the second gauge condition
$\epsilon\cdot v=0$, which we can do since the photon is on
shell.  

\begin{figure}
\centerline{
\epsfysize 3in
\epsfbox{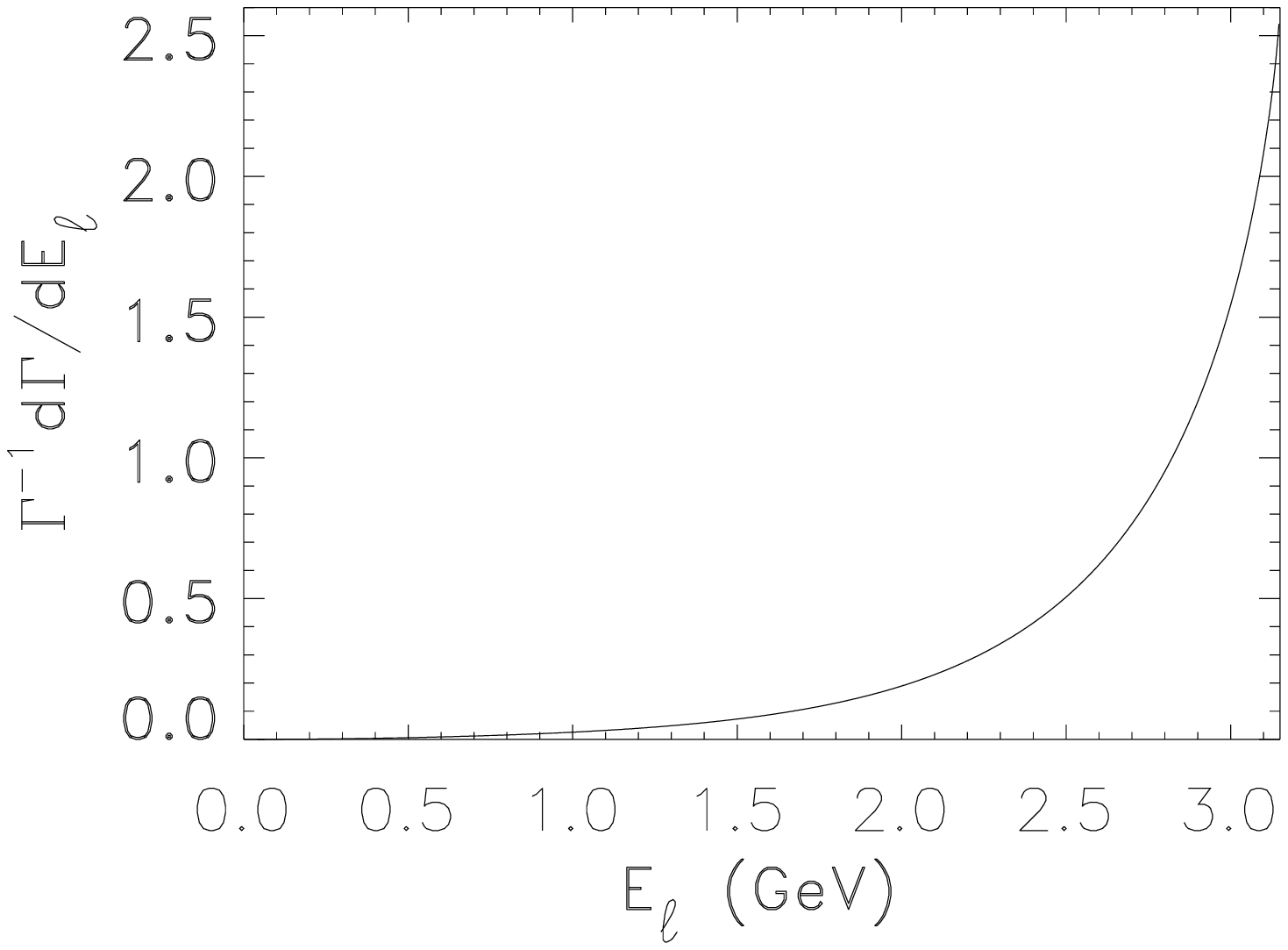}}
\vskip0.5cm
\caption{Lepton energy spectrum in $B_c \to l \nu \gamma$ 
decay.}
\label{leptonspectrum}
\end{figure}

This is the leading result in the nonrelativistic expansion. 
It differs from existing calculations in two
respects~\cite{Cha97,Ali98}.  First, it is model-independent. 
Second, radiation from the lepton leg is included, which is
required for gauge invariance of the amplitude.  At this
point, it is straightforward to derive the total rate and the
photon and lepton energy distributions.   Clearly, they depend
only on the one parameter $f_{B_c}$. Since the same
parameter enters the expression for the rate of
$B_c\to l\nu$, we will normalize the rate to that for 
$B_c\to l\nu$, and keep the dependence on the fermion charges
explicit so that the various contributions can be examined. 
Summing over the photon and lepton polarizations and
integrating over the phase space, we find the decay rate
\begin{equation}\label{rate}
  {\Gamma(B_c\to l\nu\gamma)\over\Gamma(B_c\to l\nu)}=
  \left[{\alpha\over4\pi}\,{m_{B_c}^2\over m_l^2}\right]
  m_{B_c}^2\left(\frac19{Q_b^2\over m_b^2}
  +\frac19{Q_c^2\over m_c^2}-\frac29{Q_bQ_l\over m_bm_{B_c}}
  +\frac29{Q_cQ_l\over m_cm_{B_c}}
  +\frac29{Q_l^2\over m_{B_c}^2}\right).
\end{equation} We see that there is no interference between
the photon emitted from the charm and bottom legs.  This is a
consequence of the anticorrelation of the spins of the two
quarks in the pseudoscalar $B_c$.  The dimensionless
coefficient 
$r_l=(\alpha/\pi)(m_{B_c}^2/m_l^2)$ is $r_\mu=2.1$ for $l=\mu$
and $r_e=8.8\times 10^5$ for $l=e$.  Taking $m_b=4.8\,$GeV,
$m_c=1.5\,$GeV and $m_{B_c}=6.3\,$GeV, we find
\begin{eqnarray}\label{totalrate}
  {\Gamma(B_c\to l\nu\gamma)\over\Gamma(B_c\to l\nu)}&=&
  r_l\left(0.19\,Q_b^2+1.96\,Q_c^2-0.29\,Q_bQ_l+0.93\,Q_cQ_l
  +0.22\,Q_l^2\right)\nonumber\\
  &=&r_l\left(0.02+0.87-0.10-0.62+0.22\right)\nonumber\\
  &=&0.40\,r_l\,.
\end{eqnarray}
\begin{figure}
\centerline{
\epsfysize 3in
\epsfbox{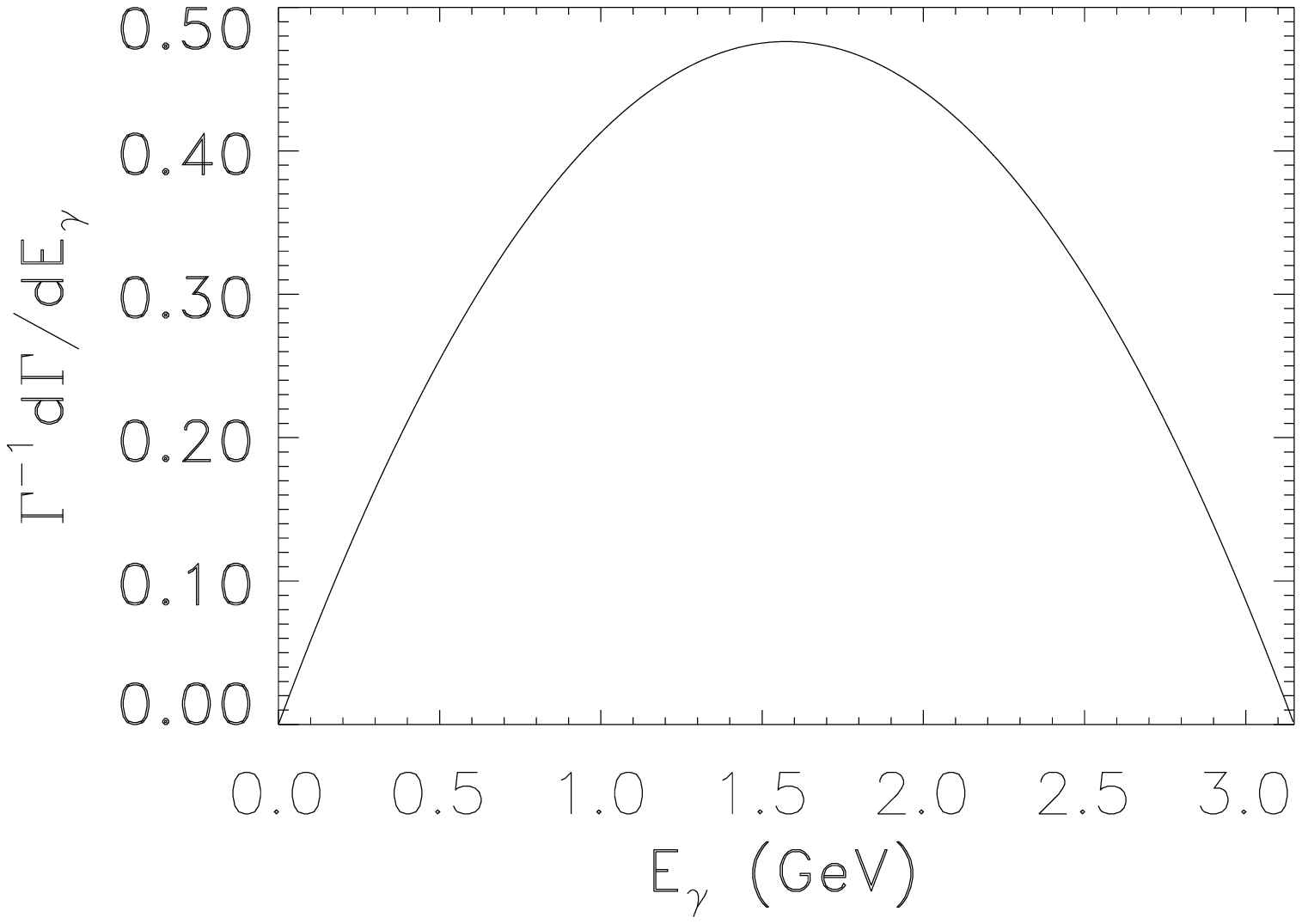}}
\vskip0.5cm
\caption{Photon energy spectrum in $B_c \to l \nu \gamma$ 
decay.}
\label{photonspectrum}
\end{figure}%
Hence the rate for $B_c\to\mu\nu\gamma$ is
approximately 80\% of that for $B_c\to\mu\nu$, while $B_c\to
e\nu\gamma$ dominates $B_c\to e\nu$.  The doubly differential
spectrum in $x=2E_\gamma/m_{B_c}$ and
$y=2E_l/m_{B_c}$ is
\begin{eqnarray}\label{doublediff}
  {1\over\Gamma}{{\rm d}\Gamma\over{\rm d}x{\rm d}y}=&&
  18(1-x)x^{-2}\Big(\mu_c^2(1-6\mu_b)
  +2\mu_b^2(2-6\mu_c+9\mu_c^2)\Big)^{-1}
  \Big[\mu_c^2(1-6\mu_b)(1-y)^2
  \nonumber\\
  &&\qquad\mbox{}+4\mu_b^2(1-3\mu_c)
  (1-x-y)^2+9\mu_b^2\mu_c^2(2-2x+x^2-4y+2xy+2y^2\Big]\,,
\end{eqnarray} where $\mu_b=m_b/m_{B_c}$ and
$\mu_c=m_c/m_{B_c}$.  The normalized photon energy spectrum
follows the simple shape
\begin{equation}\label{phospec}
  {1\over\Gamma}{{\rm d}\Gamma\over{\rm d}x}=6x(1-x)\,.
\end{equation} 
The photon energy spectrum is shown in
Fig.~\ref{photonspectrum}.  Note that there is no soft
divergence in the limit $E_\gamma\to0$.  This is because the
helicity suppression for $B _c$ decay can only be lifted by
recoil effects or a heavy quark spin flip from a magnetic
transition, both of which vanish with $E_\gamma$.  The lepton
energy spectrum, whose analytic form is
\begin{eqnarray}\label{lepspec}
  {1\over\Gamma}{{\rm d}\Gamma\over{\rm d}y}=&&
  9\Big(\mu_c^2(1-6\mu_b)
  +2\mu_b^2(2-6\mu_c+9\mu_c^2)\Big)^{-1}\nonumber\\
  &&\times
  \Big[2\mu_c^2y(1-6\mu_b)(1-y)
  +\mu_b^2y\Big(24+72\mu_c^2-63\mu_c^2y-20y
  -72\mu_c+60\mu_cy\Big)\\
  &&\quad
  \mbox{}-2(1-y)\ln(1-y)\Big(2\mu_b^2\big((6\mu_c-2)(3-y) 
  -9\mu_c^2(2-y)\big)+\mu_c^2(6\mu_b-1)(1-y)\Big)\nonumber
  \Big]\,,
\end{eqnarray} is shown in Fig.~\ref{leptonspectrum}.  Unlike
the photon energy spectrum, the lepton energy spectrum depends
on the mass ratios
$\mu_c$ and $\mu_b$.
Finally, we compute the branching ratio for
$B_c\to l\nu\gamma$.  Taking for the
$B_c$ lifetime the CDF central value 
$\tau (B_c) = 0.46\,$ps~\cite{CDFBc}, and using the estimate
$f_{B_c}=375\,$MeV~\cite{BrF95}, we obtain
\begin{equation}
  {\cal B}(B_c \to e\nu \gamma)\simeq {\cal B} (B_c \to 
  \mu\nu\gamma) \simeq 4.4 \times 10^{-5}\,.
\end{equation}  
It is certainly a challenge to observe such rare processes,
even with the large $B_c$ samples which one hopes will be
available at future experiments such as BTeV and LHC.

This result is similar to what one would find using a
constituent quark model.  However, in a number of respects
it goes beyond the quark model framework.  First, the result
holds even when the photon is hard and the virtual $c$ or
$b$ quark is far from its mass shell.  Second, the it
is rigorously true in the limit $m_c,m_b\to\infty$.  Third,
it is possible to improve systematically the accuracy of
this result by including both QCD radiative corrections and
higher dimension operators in NRQCD.  The QCD calculation
provides insight into both the success and the limitations of
the quark model picture.

Note that for very soft photons, the shape of the spectrum
is dominated by the nearest $\bar bc$ resonance, the $B_c^*$,
and the parton-hadron duality which underlies the operator
product expansion breaks down.  (Since the splitting between
the $B_c$ and the $B_c^*$ is caused by a hyperfine
interaction, it is much smaller than the splittings between
the $B_c$ and all other $B_c^{**}$ excitations.)  This corner
of phase space will be examined in more detail in Section IV. 
At larger photon energies, all virtual
$B_c^{**}$ states contribute equally to the shape, but the
integration over the momenta of the leptons smears the effect
of the resonances into a smooth result.  By the usual
application of global parton-hadron duality, the smeared
spectrum should be reproduced by perturbative QCD
(supplemented by a hierarchy of nonperturbative contributions
from operators of higher dimension).  Note that the kinematics
forbids the production of on-shell $B_c^{**}$ intermediate
states.  Nor do on-shell $\bar cc$ bound states contribute
to this exclusive process via the diagram in
Fig.~\ref{QCDgraphs}a, since the physical photon has $k^2=0$.

\section{The NRQCD expansion for the $B_c$}

We now turn to the nonrelativistic limit of the QCD answer,
matching onto a tower of NRQCD operators.  The purpose is to
identify the leading contribution to the $B_c$ matrix elements
which contribute to $B_c\to\gamma l\nu$.  (The result, which
turns out to be simple, was already anticipated in the previous
section.)  We will write our expansion in terms of effective
fields
$\Psi(x)$, which are related to the usual QCD fields
$Q(x)$ by a Foldy-Wouthuysen transformation,
\begin{equation}\label{foldy}
  Q(x)=\exp\{i\rlap{D}{\,/}_\perp/2m_Q\}\times\Psi(x)=
  \exp\{i\overrightarrow D\cdot\overrightarrow\gamma/2m_Q\}
  \times\Psi(x)\,,
\end{equation}
where $D^\mu_\perp=D^\mu-v^\mu v\cdot D$. In the rest frame of
the $B_c$ and in the Dirac representation,
$\Psi$ may be decomposed as
\begin{equation}
  \Psi=\pmatrix{\psi\cr \chi}\,,
\end{equation} where $\psi$ and $\chi$ are the two-component
quark and antiquark fields, which the Foldy-Wouthuysen
transformation disentangles to order $1/m_Q$.  The field
$\psi$ annihilates quark states, while $\chi$ creates
antiquark states.  Note that the projection operators
$P_\pm=(1\pm\rlap/v)/2$, where $v^\mu=(1,0,0,0)$ is the
four-velocity of the $B_c$, project out separately the
$\psi$ and $\chi$ parts of $\Psi$.  It is often  convenient to
use this covariant (and representation-independent) form of the
projection.

The transformation (\ref{foldy}) can be used to rewrite QCD
operators in terms of the effective fields $\Psi$.  This is
one step in the matching of QCD onto NRQCD.  For example
consider the expansion of the weak current
$\bar b\Gamma^\mu c=\bar b\gamma^\mu(1-\gamma^5)c$ to order
$1/m_{c,b}$.  In the four-component notation, we find
\begin{equation}\label{fourexpand}
  \bar b\Gamma^\mu c\to \bar
  \Psi_b\Gamma^\mu\Psi_c
  +{1\over2m_c}\Psi_b\Gamma^\mu i
  \rlap{D}{\,/}_\perp\Psi_c
  -{1\over2m_b}\Psi_b i\overleftarrow{\rlap{D}{\,/}}_\perp
  \Gamma^\mu\Psi_c\,.
\end{equation} In the two-component notation, this becomes
\begin{eqnarray} \label{current}
  \bar b\gamma^\mu(1-\gamma^5)c\to&& -g^{\mu 0} \left[
  \chi^\dagger_b \psi_c +{1\over2m_b} \chi^\dagger_b
  i\overleftarrow{D}^j \sigma^j \psi_c +
  {1\over2m_c}\chi^\dagger_b\sigma^j iD^j \psi_c\right]
  \nonumber \\
  &&- g^{\mu j} \left[\chi^\dagger_b \sigma^j \psi_c
  +{1\over2m_b} \chi^\dagger_b i\overleftarrow{D}^j \psi_c
  + {1\over2m_c}\chi^\dagger_b iD^j \psi_c\right.\\
  &&\left.\qquad\qquad -{1\over2m_b} i \epsilon^{jkl}
  \chi^\dagger_b i\overleftarrow{D}^k \sigma^l \psi_c
  +{1\over2m_c}
  i\epsilon^{jkl} \chi^\dagger_b iD^k \sigma^l \psi_c
  \right], \nonumber
\end{eqnarray} which is somewhat more cumbersome but also more
explicit.  These leading terms in the expansion in $1/m_{b,c}$
are also the leading terms in the velocity expansion; the
first omitted terms are of order ${\bf v}^2$.  A more complete
calculation would include radiative corrections as well, in
which case the various operators which appear on the right
hand sides of Eqs.~(\ref{fourexpand}) and (\ref{current})
would develop a dependence on the renormalization scale.  Note
that the covariant derivative $D_\mu$ includes both gluon and
photon gauge fields.

The effective fields $\Psi$ realize explicitly the heavy quark
spin symmetry which emerges as the magnetic interactions of $Q$
decouple in the limit $m_Q\to\infty$.  An immediate implication
of this symmetry in NRQCD is that the $B_c$ and the $B_c^*$ are
degenerate.  Another is the equality of their decay constants,
\begin{equation} \label{deccon}
  i m_{B_c} f_{B_c}=\langle 0 |\chi_b^\dagger \psi_c | B_c
  \rangle =
  \eta^i \langle 0 |\chi_b^\dagger \sigma_i \psi_c | B_c^*
  \rangle + O({\bf v}^2).
\end{equation} A particularly nice realization of this
symmetry is available in terms of the four-component effective
fields.  One can assemble the $B_c$ and the $B_c^*$ into a
single ``superfield''~\cite{LMS,Fa90},
\begin{equation}
  H_{B_c}={1+\rlap/v\over2}\left[B_c^{*\mu}\gamma_\mu
  -B_c\gamma^5\right].
\end{equation} Matrix elements which conserve the spin
symmetry then take the form
\begin{eqnarray}\label{traces}
  \langle0|\overline\Psi_b\Gamma\Psi_c|B_c^{(*)}\rangle&=&
  \frac{i}{2}f_{B_c}m_{B_c}{\rm Tr}\left[\Gamma
  H_{B_c}\right],\nonumber\\
  \langle
  B_c^{(*)}|\overline\Psi_c\Gamma\Psi_c|B_c^{(*)}\rangle
  &=&-m_{B_c}{\rm Tr}\left[\overline H_{B_c}
  \Gamma H_{B_c}\right],\nonumber\\
  \langle
  B_c^{(*)}|\overline\Psi_b\Gamma\Psi_b|B_c^{(*)}\rangle
  &=&m_{B_c}{\rm Tr}\left[H_{B_c}
  \Gamma\,\overline  H_{B_c}\right].
\end{eqnarray} The leading contribution to the mass difference
between the
$B_c$ and the $B_c^*$ comes from the chromomagnetic operator
\begin{equation}
  O_2={g_s\over4m_c}\overline\Psi_c\sigma^{\alpha\beta}
  G_{\alpha\beta}\Psi_c+
 {g_s\over4m_b}\overline\Psi_b\sigma^{\alpha\beta}
  G_{\alpha\beta}\Psi_b\,.
\end{equation} The mass difference $\Delta=m_{B_c^*}-m_{B_c}$
may be written in terms of the matrix elements of $O_2$,
\begin{equation}
  \Delta = \langle B_c^*|O_2|B_c^*\rangle -
  \langle B_c|O_2|B_c\rangle .
\end{equation} This relation will be useful later, when we
consider the effect of the $B_c^*$ pole.

\subsection{NRQCD calculation and matching}

Due to the simplicity of the final state, we can match the
currents of QCD directly onto NRQCD, rather than the decay
rates.  For example, Lorentz invariance restricts the form of
the NRQCD current which annihilates the $B_c$ to
\begin{equation}\label{nrqcdexpand}
  \langle 0|j^\mu |B_c \rangle = -g^{\mu 0} \left\{ C_1
  \langle 0
  | \chi^\dagger_b \psi_c |B_c \rangle +
  \frac{C_2}{m_{red}^2}
  \langle 0 | \chi^\dagger_b\,
  \vec{D}^2\psi_c|B_c\rangle+\ldots
  \right \},
\end{equation}
where $m_{red}=m_b m_c/(m_b + m_c) $ is the reduced mass of
the $\bar bc$ pair.  At leading order, this reduces to the
matrix element of a single universal NRQCD operator which can
be  either fixed by other experimental measurements or related
to the $B_c$ wave function at the origin~\cite{EiQu95}.

We would like to generalize the effective Lagrangian of  NRQCD
to include the effects of the weak interactions. This can be
achieved by introducing external sources. The effective NRQCD
Lagrangian can be separated into two sectors representing
flavor conserving (${\cal L}_{FCon}$) and flavor changing
(${\cal L}_{FCh}$)  interactions, as well as a source part
${\cal L}_S$,
\begin{equation} \label{slag}
  {\cal L}_{NRQCD} = 
  {\cal L}_{FCon} + {\cal L}_{FCh} + {\cal L}_S\,.
\end{equation} The sources represent external perturbatively
interacting fields. In addition to the electromagnetic  field,
in this calculation it is convenient to treat the leptonic
current as an external source field responsible for the flavor
changing interactions.  Each part of this Lagrangian can be
organized by the velocity expansion.  At leading order, the
flavor conserving part can be written
\begin{eqnarray} \label{LCon}
  {\cal L}_{FCon}^L &=& \chi^\dagger_\alpha 
  \left(i D_t - {\vec{D}^2}/{2 m_\alpha} \right)
  \chi_\alpha + 
  \psi^\dagger_\alpha 
  \left(i D_t + {\vec{D}^2}/{2 m_\alpha} \right)
  \psi_\alpha \nonumber \\
  &&\qquad+\frac{a_2^\alpha}{2 m_\alpha} \left (
  \psi^\dagger_\alpha 
  \vec{\sigma} \cdot \vec{v}_2 \psi_\alpha
  - \chi^\dagger_\alpha \vec{\sigma} \cdot \vec{v}_2
  \chi_\alpha \right ),
\end{eqnarray} where the covariant derivatives include
external scalar $s$, $D_t= \partial_t + a_0^\alpha s$ and
vector $\vec{v}_1$, $\vec{D}= \vec{\partial} + i a_1^\alpha
\vec{v}_1$  sources, and summation over quark flavors
$\alpha=\{c,b\}$ is understood. Here
$a_i^\alpha$ represent the coefficients of NRQCD operators,
which are normalized such that $a_i^\alpha = 1 +
O(\alpha_s)$.  Note that in this notation, the field $\chi$
creates an antiquark, so if the Lagrangian (\ref{LCon}) were
normal ordered then the kinetic term for the antiquark field
would have the opposite sign.  At leading order, the flavor
changing part of Eq.~(\ref{slag}) reads
\begin{eqnarray} \label{LCh}
  {\cal L}_{FCh}^L = - a_3 \chi^\dagger_\alpha 
  \psi_\beta S_1 -
  a_4 \chi^\dagger_\alpha 
  \vec{\sigma} \cdot \vec{V}_1
  \psi_\beta~ +{\rm \ h.c.}\,,
\end{eqnarray} 
where $S_1$ and $\vec{V}_1$ represent
flavor-changing external  sources. It is important to remember
that the power counting rules for the Lagrangian with external
sources are  different from the usual NRQCD Lagrangian, in
that one counts only the powers of quark fields. Therefore,
the next-to-leading Lagrangian contains covariant  derivative
insertions into the leading order Lagrangian,
\begin{equation} \label{NLCon}
  {\cal L}_{FCon}^{NL} = 
  \frac{b_1^\alpha}{8 m_\alpha^2} \left[
  \chi^\dagger_\alpha 
  D^i \chi_\alpha +
  \psi^\dagger_\alpha 
  D^i \psi_\alpha \right] v^i_3
  +\frac{b_2^\alpha}{8 m_\alpha^2} \left[
  \chi^\dagger_\alpha 
  \left (
  \vec{D} \times \vec{\sigma} \right )^i \chi_\alpha
  + \psi^\dagger_\alpha \left(
  \vec{D} \times \vec{\sigma} \right)^i \psi_\alpha
  \right ] v^i_3\,.
\end{equation} 
For the external electromagnetic field one
identifies
$s_1$ and $\vec{v}_1$ with the external electromagnetic
potentials, while $\vec{v}_2$ and $\vec{v}_3$ represent
external magnetic and electric fields,
$v_2^i= e Q_\alpha {\bf B}^i =  {1\over2} e Q_\alpha
\epsilon^{ijk} F^{jk}$ and $v_3^i= e Q_\alpha {\bf E}^i = e
Q_\alpha F^{0i}$. At next-to-leading order, the flavor
changing Lagrangian is
\begin{eqnarray} \label{NLCh}
  {\cal L}_{FCh}^{NL} &=& 
  -\frac{b_3}{2} \chi^\dagger_\alpha 
  \left( \frac{i\overleftarrow{D}^i}{m_\alpha} +
  \frac{iD^i}{m_\beta} \right) \sigma^i \psi_\beta S_2 -
  \frac{b_4}{2} \chi^\dagger_\alpha 
  \left( \frac{i\overleftarrow{D}^i}{m_\alpha}
  +\frac{iD^i}{m_\beta} \right) \psi_\beta V_2^i 
  \nonumber \\
  &&\qquad+ \frac{b_5}{2} \epsilon^{ikl} \chi^\dagger_\alpha 
  \left( \frac{\overleftarrow{D}^i}{m_\alpha} -
  \frac{D^i}{m_\beta} \right ) \sigma^k \psi_\beta V_2^l +
  {\rm \ h.c.}\,. 
\end{eqnarray} 
We will identify $S_i$ and $\vec{V}_i$ with the
time and space components of the leptonic current  
$L_\mu =V_{cb} (G_F/\sqrt{2})  \bar u \gamma_\mu (1-\gamma_5)
v= V_{cb} (G_F/\sqrt{2}) l_\mu$.  As defined in
Eqs.~(\ref{NLCon}) and (\ref{NLCh}), the coefficients satisfy
$b_i^\alpha=1+O(\alpha_s)$.

A set of NRQCD Feynman rules follows directly from this
Lagrangian and can be used to calculate the set of diagrams of
Fig.~\ref{QCDgraphs}. The calculation of diagram
Fig.~\ref{QCDgraphs}c is identical to the one in QCD and yields
\begin{equation} \label{c}
  {\cal A}_c = a_3 V_{cb} \frac{G_F eQ_l}{2\sqrt{2} (p_l k)} 
  \left\{ 2 v^\mu p_l^\nu + (k v) g^{\mu \nu}- k^\mu v^\nu  - 
  i \epsilon^{\mu \alpha \nu \beta} v_\alpha k_\beta 
  \right \} \epsilon^*_\nu l_\mu
  \langle 0 | \chi_b^\dagger \psi_c | \bar b c (^1S_0)\rangle,
\end{equation} The amplitudes of diagrams
Fig.~\ref{QCDgraphs}a and Fig.~\ref{QCDgraphs}b can be
computed using Feynman rules derived from  Eqs.~(\ref{LCon})
and (\ref{LCh}),
\begin{eqnarray} \label{ab}
  {\cal A}_{a,b} &=& -i a_2 a_4 V_{cb}\frac{G_F}{\sqrt{2}}
  \epsilon^{ijk} l^i k^j \epsilon^{*k} 
  \langle 0 | \chi^\dagger_b \psi_c |
  \bar b c (^1S_0) \rangle
  \nonumber\\ &&\qquad\qquad\times
  \left [
  - \frac{e Q_b}{2 m_b (p^0_1-\vec{p}_1^{\: 2}/2m_b)} + 
  \frac{e Q_c}{2 m_c (p^0_2-{\vec{p}}_2^{\: 2}/2m_c)} \right],
\end{eqnarray} where $p^0_1=E_b+E_\gamma,~\vec{p}_1 =
\vec{p}_b + \vec{k}$,
$p^0_2=E_c-E_\gamma,~\vec{p}_2 = \vec{p}_c - \vec{k}$.

To compare Eq.~(\ref{ab}) to the corresponding expression in
full QCD, it is convenient to adopt a covariant notation and
choose the gauge\footnote{Clearly,  
the matching procedure can be performed in any gauge; in
NRQCD, this gauge choice is most convenient and quite common.}
$\epsilon\cdot v=0$. Then it can be rewritten as
\begin{eqnarray} \label{abfinal}
  {\cal A}_{a,b} = i a_2 a_4 V_{cb}\frac{G_F}{\sqrt{2}}
  {1\over kv}\left [
  \frac{e Q_b}{2 m_b} + \frac{e Q_c}{2 m_c} \right]
  \epsilon^{\mu \alpha \nu \beta} l_\mu
  \epsilon^{*}_\nu  v_\alpha  k_\beta
  \langle 0 | \chi^\dagger_b \psi_c |
  \bar b c (^1S_0) \rangle.
\end{eqnarray} In addition, there is also a local interaction
coming from a $b_4$ term of the flavor changing part of the
Lagrangian (\ref{NLCh}),
\begin{eqnarray} 
  {\cal A}_s = -
  b_4 V_{cb} \frac{G_F}{\sqrt{2}} \frac{i}{2} 
  g^{\mu i} l_\mu \langle \gamma |
  \chi^\dagger_b \left(\frac{\overleftarrow{D}_i}{m_b} + 
  \frac{D_i}{m_c} \right)
  \psi_c |\bar b c (^1S_0) \rangle
\end{eqnarray}  This contribution involves operators with
covariant derivatives. Recalling that  covariant derivatives
contain  electromagnetic source fields whose couplings can be
treated perturbatively, we obtain
\begin{equation} \label{seagull}
  {\cal A}_s =
  -b_4 V_{cb} \frac{G_F}{2\sqrt{2}}
  \left(\frac{eQ_c}{m_c} -\frac{eQ_b}{m_b} \right)
  \left [ g^{\mu \nu} - v^\mu v^\nu \right ]
  \epsilon^*_\nu l_\mu
  \langle 0 | \chi^\dagger_b \psi_c |
  \bar b c (^1S_0) \rangle.
\end{equation} Note that this term has no dependence on the
photon energy. Adding the
contributions (\ref{c}), (\ref{abfinal}) and (\ref{seagull})
and comparing with the QCD result (\ref{qcd1}), we reproduce
at leading order the anticipated matching conditions
$b_i^\alpha=a_i^\alpha=1$.

We hasten to add a remark here.  At leading order, the only
contribution that survives in the  denominators of
Eq.~(\ref{ab}) is $E_\gamma \equiv kv$, matching the propagator
of full QCD.  However, in NRQCD this calculation is only good
for $E_\gamma\ll m_{c,b}$, where the heavy quarks are close to
their mass shell.  There is a substantial portion of the final
state phase space in which the propagating heavy quark is
relativistic and therefore not described by the formalism of
NRQCD at  the leading order.  In a formal sense, the sum of an
infinite number of NRQCD operators is required to fully
describe the propagation of the relativistic quark.  We can
separate the region in which NRQCD is valid from the one in
which it is not by introducing a factorization scale
$\mu\ll m_{c,b}$.  Then the amplitude (\ref{abfinal})  is
generated by the NRQCD Feynman diagrams in
Fig.~\ref{QCDgraphs} only for $E_\gamma<\mu$.   For the region
$E_\gamma>\mu$, we must add a local  ``operator'' to the
Lagrangian,
\begin{equation}\label{newop}
  -V_{cb}{G_F\over\sqrt2}{1\over kv}\left({eQ_b\over2m_b}
  +{eQ_b\over2m_b}\right)\epsilon^{ijk} l^i k^j \epsilon^{*k} 
  \,\chi^\dagger_b \psi_c 
\end{equation} to reproduce NRQCD amplitude in this kinematic
region.  This object is an operator in the NRQCD quark fields,
but a c-number with respect to the external electromagnetic
sources.  Note that its inclusion is required by the matching
conditions of QCD onto NRQCD, to account for the situation
in which the intermediate heavy quark is far from its mass
shell~\cite{bbl}.  This is entirely in accord with what one
usually finds upon applying the operator product expansion,
that highly virtual intermediate states generate operators of
higher dimension in the low energy theory.  Note that
since the this new object is to be included only in the case
of large photon energy $E_\gamma>\mu$, it could equally well be
written in terms of a local operator built out of the photon
field strength $F_{\mu\nu}$.  We have chosen to write
Eq.~(\ref{newop}) in this form so that it may be compared
transparently to the amplitudes which contribute at small
$E_\gamma$.

It is now clear that the leading contribution to the decay
$B_c\to\gamma l\nu$ indeed comes from the dimension three
matrix element
$\langle 0|\chi_b^\dagger\psi_c|B_c\rangle=im_{B_c}f_{B_c}$,
as anticipated in the previous section.  This entirely
unsurprising result has now been placed on firm theoretical
footing, as the leading term in a systematic expansion. 
Furthermore, we are in a position to estimate the size of the
most important corrections.  From the expansion
(\ref{nrqcdexpand}), we see that there is a relativistic
correction to $\langle 0|\chi_b^\dagger\psi_c|B_c\rangle$ of
relative order ${\bf v}^2$ by NRQCD
power counting~\cite{bbl}.  This is a contribution to
$f_{B_c}$, however, which cancels in the ratio
(\ref{totalrate}).  The leading correction which is particular
to $B_c\to\gamma l\nu$ comes from the dimension five
operator $\langle0|\chi_b^\dagger\, g{\vec\sigma}\cdot\vec{\bf
B}\psi_c|B_c\rangle/m_c^2$, generated by attaching a soft
gluon to the $c$ quark line in Fig.~\ref{QCDgraphs}a.  This
matrix element is of relative order ${\bf v}^4\sim5\%$.  In NRQCD, matrix elements such as
$\langle 0\,|\chi^\dagger_b\,\vec{D}^2\psi_c\,|B_c\rangle$ and 
$\langle0|\chi_b^\dagger\, g{\vec\sigma}\cdot\vec{\bf
B}\,\psi_c|B_c\rangle$ are universal quantities which
can, in principle, be extracted from other $B_c$ decays. 
Note that the relativistic contributions are formally smaller
than QCD radiative corrections of order $\alpha_s$, which we
have not included.\footnote{The leading radiative correction
to $B_c\to l\nu$ was computed in Ref.~\cite{BrF96}.}

\section{Very soft photons and the $B_c^*$ pole}

For weak radiative $B_c$ decays, there is the possibility that
the outgoing photon is very soft, so much so that there is a
large time separation between the event where the photon
is emitted and the event where quarks annihilate to
the lepton pair.  In this case, the physics is not adequately
described by (even nonrelativistic) quark fields, and
consideration of hadronic intermediate states is necessary. 
Let us study this part of the spectrum more carefully.  The
treatment in this section is similar to that of
Ref.~\cite{Bur94} for the decay $B\to\pi l\nu$.

\begin{figure}
\centerline{
\epsfxsize 6in
\epsfbox{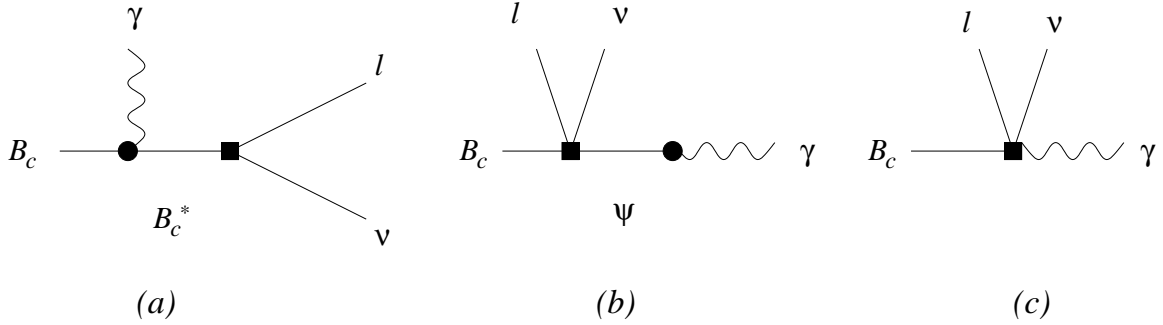}}
\vskip0.5cm
\caption{Pole structure in $B_c \to l \nu \gamma$, at leading
order.}
\label{Bcpole}
\end{figure}

The amplitude ${\cal A}(B_c\to l\nu\gamma)$ is a second-order
process involving contributions both from the electromagnetic
part of the Hamiltonian $H_{em}$ and from the weak part
$H_w$.  For a sufficiently soft photon, in which case the
recoil of the hadrons can be neglected, the amplitude is
\begin{eqnarray} \label{specampl0} 
  {\cal A}(B_c \to  l \nu \gamma) & = &
  i\int d^4 x\, e^{ikx}\,\langle l \nu\gamma |\, T\left\{H_{el}
  (x), H_w (0) \right\} |B_c \rangle \nonumber \\  
  & = & \sum_M
  \langle l\nu |H_w|M\rangle
  \frac{1}{\Delta E_M} 
  \langle M \gamma |H_{el}|B_c\rangle ~+~\mbox{local terms},
\end{eqnarray}  where $\Delta E_M$ is the energy
difference between the $B_c$ and the $M\gamma$ intermediate
state, and $M=B_c,B_c^*,B_c^{**},\ldots$ represents any meson
of the $B_c$ family.  Heavy quark spin symmetry implies
that the mesons $M$ are found in degenerate pairs, of which
the $B_c$ and $B_c^*$ comprise the lightest.  Heavy quark
symmetry also implies that in the limit $E_\gamma\to0$, the
matrix element $\langle M \gamma |H_{el}|B_c\rangle$ vanishes
except when $M$ is a member of the ground state doublet
$(B_c,B_c^*)$.  In this limit, this is a ``zero-recoil''
transition, such as at the $w=1$ point in semileptonic $B$
decays~\cite{IW}.  Therefore, for very small
$E_\gamma$  the sum in Eq.~(\ref{specampl0}) is dominated by
the contributions from $B_c$ and $B_c^*$.  Since the weak
$B_c\to l\nu$ transition is helicity suppressed for
$m_l=0$, only the vector $B_c^*$ state survives the sum. 
This situation is represented in Fig.~\ref{Bcpole}a.  

We will neglect contributions from the graphs of
Fig.~\ref{Bcpole}b, corresponding to intermediate $\psi$
states. These graphs are proportional to $g_{\gamma
\psi}/m_{\psi}^2$  and are formally suppressed compared to the
$B_c^*$ pole.  Moreover, since the $\psi$ states are far off
shell, these  contributions may be systematically accounted
for by including  a set of higher-order NRQCD operators.  Note
that the situation is completely different in $B_u \to l \nu
\gamma$ decays, where the analogous contribution is from an
intermediate $\rho$ meson and cannot be calculated in a model
independent fashion in HQET, thus providing an intrinsic
uncertainty. There are also local terms arising
from the graphs like Fig.~\ref{Bcpole}c. Concentrating on the
``pole'' part,
\begin{eqnarray} \label{specampl} 
  {\cal A}(B_c \to  l \nu \gamma)  =  \sum_M
  i \langle l\nu |H_w|B_c^*\rangle
  \frac{i}{2m_{B_c}v\cdot (-k)} 
  \langle B_c^* \gamma |H_{el}|B_c\rangle+\ldots\,.
\end{eqnarray}  Clearly, the position of the pole in
Eq.~(\ref{specampl}), although consistent with a heavy quark
symmetry expectations, is not correctly located at $E_\gamma
=0$.  We now show how the leading chromomagnetic corrections
move the pole to the right place.

At this stage it is convenient to adopt an
HQET notation, in which the NRQCD operators are build from
four-component spinors constrained by a set of on-shell
conditions,
\begin{equation}
\rlap/{v} \Psi_c =  \Psi_c\,, \qquad \bar \Psi_b \rlap/{v} = 
-\bar \Psi_b.
\end{equation}   In this formalism, the electromagnetic part of
the Hamiltonian contains two parts, which can be identified
with the spin-conserving electric and spin-flipping  magnetic
transitions,
\begin{equation} 
  H_{em}  = \frac{eQ_c}{4m_c}
  \bar \Psi_c  {\sigma}^{\alpha \beta} F_{\alpha \beta} \Psi_c
+
  \frac{eQ_b}{4m_b}
  \bar \Psi_b  {\sigma}^{\alpha \beta} F_{\alpha \beta}
\Psi_b\,
\end{equation}  where $ F_{\alpha \beta}=\partial_{\alpha}
A_{\beta} - \partial_{\beta} A_{\alpha} $ is the
electromagnetic field strength tensor.  Clearly, only the
magnetic part of
$H_{em}$ will contribute to the $B_c^*$ intermediate
state in Eq.~(\ref{specampl}). The
$\langle B_c^* \gamma | H_{em} | B_c \rangle$ matrix
element is then
\begin{equation}
  i \langle B_c^* \gamma | H_{em} | B_c\rangle =
  -k_{\alpha}\epsilon_{\beta}^*
  \left[ \frac{eQ_c}{2m_c} \langle B_c^* | \bar \Psi_c 
  {\sigma}^{\alpha \beta} \Psi_c  |B_c\rangle +
  \frac{eQ_b}{2m_b}
  \langle B_c^* | \bar \Psi_b  {\sigma}^{\alpha \beta}
\Psi_b |
  B_c\rangle \right].
\end{equation}  These matrix elements can be calculated using
the trace formalism (\ref{traces}), yielding
\begin{equation}
  \langle B_c^* | \bar \Psi_q  {\sigma}^{\alpha \beta}
  \Psi_q 
  | B_c\rangle =
  2m_{B_c}\eta_{\mu}^* v_{\nu} \epsilon^{\mu \nu \alpha
\beta} \,,
\end{equation}  where $\eta_{\mu}$ is the polarization vector
of the
$B_c^*$ meson, and $\Psi_q$ represents either heavy
effective field.

We now use heavy quark symmetry (\ref{deccon}) to relate the
decay constant of the $B_c^*$ to that of the $B_c$,
\begin{equation}
  \langle 0 | \bar \Psi_b \gamma_{\mu} (1-\gamma_5 ) \Psi_c
  |B_c^*
  \rangle  = if_{B_c} m_{B_c} \eta_{\mu} \, .
\end{equation}  Thus in the heavy quark limit and with $m_l=0$
we obtain for the soft photon amplitude
\begin{equation} \label{singlepole}
  {\cal A}(B_c \to l\nu \gamma ) =
  -\frac{V_{cb}G_F}{\sqrt{2}} f_{B_c} m_{B_c}\,
  \frac{\mu_{B_c}}{v\cdot k}\, \epsilon^{\mu \alpha \nu \beta}
  l_{\mu} v_{\alpha}
  \epsilon_{\nu}^* k_{\beta}~+~\mbox{local term},
\end{equation}  where we have introduced for convenience the
perturbative $B_c$ magnetic moment, $\mu_{B_c} = eQ_b/2m_b +
eQ_c/2m_c$. Since in the heavy quark limit the
$B_c$ and $B_c^* $ are degenerate, the pole is at the
``wrong'' position $E_\gamma = 0$.  Next we show how inclusion
of the $1/m_{c,b}$ corrections removes the degeneracy and
shifts the pole to
$E_\gamma = - \Delta= -(m_{B_c^*}-m_{B_c})$, in the
unphysical region where it belongs.

The leading $1/m_{c,b}$ corrections come from the insertion
of the kinetic energy and chromomagnetic dipole operators,
$O_{kin}=\bar \Psi_q(iD)^2\Psi_q/2m$ and $O_{mag}=g_s\bar
\Psi_q\sigma^{\alpha\beta}G_{\alpha\beta}\Psi_q/4m$.  We use
the customary definitions
\begin{eqnarray}
  \langle B_c^{(* )} |\bar \Psi_q(iD)^2\Psi_q |B_c^{(*
)} \rangle  & = &
  2m_{B_c}\lambda_1 \nonumber \\
  \langle B_c^{(* )} |\frac{g_s}{2}\bar
\Psi_q\sigma^{\alpha\beta}G_{\alpha\beta}\Psi_q |B_c^{(* )}
\rangle & = &  2m_{B_c} d_M^{(*)} \lambda_2 \, ,
\end{eqnarray} where $d_M = 3$ and $d_M^* =-1$.  Note
that unlike in HQET, in NRQCD the matrix elements of these
operators have different velocity powers; the first operator
is there at leading order, while the second is suppressed by a
power of ${\bf v}$.

\begin{figure}
\centerline{
\epsfysize 2in
\epsfbox{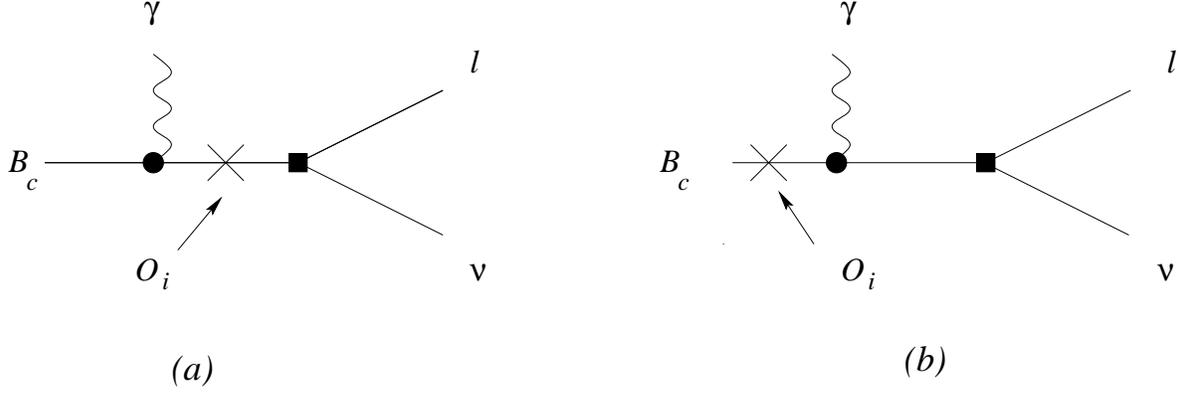}}
\vskip0.5cm
\caption{Pole structure in $B_c \to l \nu \gamma$, at
next-to-leading order.  The $O_i$ represent insertions of the
kinetic energy and chromomagnetic operators.}
\label{Bcpole2}
\end{figure}

Inserting $O_{kin}$ and $O_{mag}$ vertices on the $B_c^*$
meson line gives rise to the double pole contribution
\begin{eqnarray} \label{doublepole} 
  {\cal A}(B_c \to l \nu \gamma) &=& \langle l\nu
  |H_w|B_c^{(* )} \rangle  \langle B_c^{(* )} |O_{kin} +
  O_{mag}|B_c^{(* )} \rangle
  \langle B_c^{(* )}\gamma |H_{el} |B_c \rangle {\left(
  \frac{1}{2m_{B_c}v\cdot k} \right)}^2
  \nonumber  \\ 
  &=& - \frac{\delta^*}{2m_{B_c}{(v\cdot
  k)}^2} \langle l\nu |H_w| B_c^{(*)}\rangle \langle
  B_c^{(* )}\gamma |H_{el} |B_c \rangle \,+... \,,
\end{eqnarray}  where $\delta^* = - (\lambda_1 + d_M^*
\lambda_2)/2m_{red}$  and
$m_{red}=m_b m_c/(m_b + m_c) $. The same matrix elements
contribute to $1/m_{c,b}$ corrections to the physical meson
masses,
\begin{equation} 
  m_{B_c^{(*)}}= m_b + m_c -
  \frac{1}{2m_{red}} (\lambda_1 + d_M^{(*)} \lambda_2)\,.
\end{equation}  When $O_{kin}$ and $O_{mag}$ and inserted on
the external line as shown in Fig.~\ref{Bcpole}b, they modify
the heavy meson propagator by shifting the position of the
pole,
\begin{equation}
  \frac{i}{v \cdot k} \to \frac{i}{v \cdot k+\delta}
  =\frac{i}{v
  \cdot k} \left( 1- \frac{\delta}{v \cdot k} + ... \right) \,,
\end{equation}  where $\delta = - (\lambda_1 + d_M
\lambda_2)/2m_{red} $.   Taking this modification into account
in (\ref{singlepole}) and combinig it  with the double pole
contribution (\ref{doublepole}), one sees that  the terms
proportional to $\lambda_1$ cancel but terms proportional to
$\lambda_2$ don't, since $d_M \ne d_M^*$.  Then
the propagator in (\ref{specampl}) is replaced by 
\begin{equation}
\frac{i}{v \cdot (-k)} \to \frac{i}{v \cdot (-k)}\left[
1+\frac{\Delta}{v \cdot (-k)} \right] = -\frac{i}{v\cdot k +
\Delta} \,.
\end{equation}  Hence we obtain
\begin{equation} 
  {\cal A}(B_c \to l\nu \gamma ) =
  -\frac{G_F V_{cb}}{\sqrt{2}} f_{B_c} m_{B_c}\,
  \frac{\mu_{B_c}}{v\cdot k+\Delta}\,\epsilon^{\mu \alpha \nu
  \beta} l_{\mu} v_{\alpha}
  \epsilon_{\nu}^* k_{\beta}~+~\mbox{local},
\end{equation}  with the pole now in the correct place.  Since
the photon energy spectrum is finite, and in fact vanishes, as
$E_\gamma\to0$, the shift in the position of the pole
has little effect on the total rate.

\section{Summary}

We have performed a model-independent study of the weak
radiative $B_c$ decay in the framework of nonrelativistic
QCD.  This process  is an important competitor to the
annihilation process $B_c \to \mu\nu$, which eventually could
be used to extract the $B_c$ decay constant. We found that
the branching ratio for  $B_c \to \gamma \mu \nu_\mu$ is of the
same order of magnitude as the corresponding purely leptonic
decay, while $B_c \to \gamma e \nu_e$ completely dominates its
leptonic counterpart. We have generalized the NRQCD Lagrangian
by  introducing external sources for the electromagnetic and
weak fields.  At leading order, NRQCD gives a result similar
to what one might expect from a constituent quark model; at
higher order, we estimate that the leading nonperturbative
corrections to $\Gamma(B_c\to\gamma l\nu)/\Gamma(B_c\to
l\nu)$ are at the level of 5\%.  Finally, we showed how the
$B_c^*$ pole emerges for very soft photons.  Unfortunately,
the branching ratio for this process is small, ${\cal B} (B_c
\to l\nu\gamma)\simeq 4.4\times 10^{-5}$, and it will
certainly be a challenge to observe this process even at
future hadronic $B$ Factories.

\acknowledgments

This work was supported in part by the United States National
Science Foundation under Grant No.~PHY-9404057. A.F.~was also
supported by the United States National Science Foundation
under National Young Investigator Award No.~PHY-9457916, by
the United States Department of Energy under Outstanding
Junior Investigator Award No.~DE-FG02-94ER40869, by the Alfred
P.~Sloan Foundation, and by the Research Corporation as a
Cottrell Scholar. A.P. thanks the Fermilab Theory Group, where
part of this work was completed.

\end{document}